\documentclass[aps,twocolumn,showpacs,groupedaddress]{revtex4}

\usepackage{amsmath,graphicx}

\begin{document}

\title{Controlling quasiparticle excitations in a trapped Bose-Einstein condensate}
\author{S. J. Woo, S. Choi, and N. P. Bigelow}

\affiliation{\mbox{Department of Physics and Astronomy, and Laboratory for Laser Energetics,} University of Rochester, Rochester, NY 14627}

\date{\today}


\begin{abstract}

We describe an approach to quantum control of the quasiparticle excitations in a trapped Bose-Einstein condensate based on adiabatic and diabatic changes in the trap anisotropy. 
We describe our approach in the context of Landau-Zener transition at the avoided crossings in the quasi-particle excitation spectrum. 
We show that there can be population oscillation between different modes at the specific aspect ratios of the trapping potential at which the mode energies are almost degenerate. 
These effects may have implications in the expansion of an excited condensate as well as the dynamics of a moving condensate in an atomic wave guide with a varying width.
\end{abstract}

\maketitle

\vspace{6mm}

The trapped atomic Bose-Einstein condensate (BEC) provides a unique mesoscopic system in which the quantum mechanics of a gaseous system dominates. With the rich experimental techniques now available, we are at the point where ``quantum engineering" of the condensate is an exciting frontier. In many examples of quantum engineering, strategically tailored states are engineered by controlling the population of the excited state of the system.  For example, in cavity QED \cite{Haroche} this is achieved using the atom-field interaction to manipulate the photon state, the atom state or a combination of both.  On an atom chip\cite{Folman}, time-varying or space-varying potentials are used to control atomic motional states. Turning back to BEC, it is well recognized that many-body quantum physics, and in particular the mean field theory, accurately describes most of the properties of the BEC. In this framework, many of the dynamical properties of the condensate can be understood in terms of the quasiparticle excitations of the system. The quasiparticle excitations also provide an effective tool for probing the role of interactions and for testing theoretical models.

In this paper, we consider how changes in the shape of the harmonic trapping potential can be used to quantum control the condensate through the control of quasiparticle excitations. 
Such changes can be easily implemented, and may be encountered under different contexts such as the imperfections present in the guiding potential of an atomic wave guide for a moving condensate.  
Central to the our treatment is the control of the Landau-Zener (LZ) transitions \cite{Landau-Zener} between different types of quasiparticles.
The LZ transition is observed in various systems, from atomic and molecular physics \cite{Niu}, to solid state physics through to mesoscopic systems \cite{Sinitsyn}. However, in general, the theory only applies well to linear systems.  By contrast, the successful application of LZ theory to a BEC\cite{Ishkhanyan-Jona-Lasinio} requires more care given the intrinsic nonlinearity of the mean field that describes the condensate.

We begin with a brief description of the hydrodynamic modes in an anisotropic trapping potential based on the observed structure of the quasiparticles.
We then examine the LZ transition qualitatively and quantitatively. 
We additionally find that population oscillations between different types of excitations around the avoided level crossing is possible.
Methods to control the normal modes are discussed.

To describe the BEC ground state, we consider a pancake-shape trap geometry with a strong confinement along the $z$-axis \cite{pancake}.
A two-dimensional (2D) time-independent Gross-Pitaevskii equation (GPE) \cite{DalfovoReview} based on mean field theory is used:
\begin{equation}
\label{GPeqn}
\left({\cal H}+g\psi_0^*\psi_0-\mu\right)\psi_0 =0,
\end{equation}
where ${\cal H}=-(\hbar^2/2M)\nabla^2+V_{\rm tr}({\bf r})$. 
Here, $\psi_0({\bf r})\equiv\langle\hat{\psi}({\bf r})\rangle$ is defined following the standard mean field treatment, with $\hat{\psi}({\bf r})$ being the field operator satisfying the bosonic commutation relation. 
$g = 4\pi a\hbar^2/M$ is the interparticle coupling constant with the $s$-wave scattering length $a$. $\mu$ is the chemical potential, and the wavefunction $\psi_0$ is normalized to the total number of particles $N$.
For our calculation, 2D anisotropic harmonic potential $V_{\rm tr} = M(\omega_x^2x^2+\omega_y^2y^2)/2$ is adopted and an effective $g_{\rm 2D}=g(m\omega_z/2\pi\hbar)^{1/2}$ is used \cite{CastinDum}.
Coupled Bogoliubov-de Gennes (BdG) equations \cite{DalfovoReview} have been solved numerically to calculate low-lying energy spectrum at various aspect ratios, $\lambda \equiv \omega_y/\omega_x$:
\begin{eqnarray}
\nonumber
\left({\cal H}+2g\psi_0^*\psi_0 -\mu\right)u_j-g\psi_0^2v_j&=&\hbar\omega_ju_j \\
[-.2cm]
\label{bogol_r11} \\
[-.2cm]
\nonumber
\left({\cal H}+2g\psi_0^*\psi_0 -\mu\right)v_j-g\psi_0^{*2}u_j&=&-\hbar\omega_jv_j,
\end{eqnarray}
where $u_j$, $v_j$ and $\omega_j$ represent the wavefunctions and eigenenergies for the $j$th quasiparticle excitation.
Time-dependent GPE \cite{DalfovoReview} has been used for the examination of dynamics with varying parameters.
All values in this paper are based on the dimensions of trap potential along the $x$ axis: $\sqrt{\hbar/M\omega_x}$ for the length, and $\omega_x^{-1}$ for time.

\begin{figure}[ht]
\centerline{\includegraphics*[height=8cm]{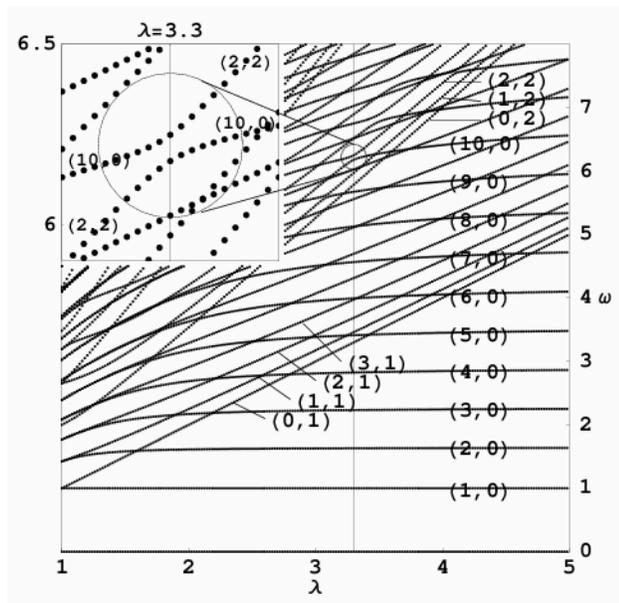}}
\caption{Energy spectrum of hydrodynamic modes of BEC with respect to the aspect ratio $\lambda = \omega_y/\omega_x$. In the circle is the avoided crossing between $(10,0)$ and $(2,2)$ modes.}
\label{spectrum}
\end{figure}
The quasiparticle energy spectrum from solving BdG equations for different values of the aspect ratio, $1<\lambda<5$, is shown in Fig. \ref{spectrum}.
As $\lambda$ increases, the BEC undergoes an effective dimensional change from 2D to almost 1D.
On the left end of the plot, the aspect ratio of the trap is 1, for which the full spectrum had been found \cite{Woo}. 
In this case, every energy level has a degeneracy of 2 and is characterized by the angular quantum numbers $\pm m$ and a radial quantum number $n$. 
If the aspect ratio deviates from unity, this degeneracy is lifted: 
the degenerate states $(n, \pm m)$ become two non-degenerate states $(n_x=|m|-1, n_y=2n+1)$ and $(n_x=|m|, n_y=2n)$.
The new set of quantum numbers $(n_x, n_y)$ corresponds to the number of zeroes along each axis in the quasiparticle wave functions $u_j$ and $v_j$.
Examination of the time-dependent probability density $|\psi_0 + u_{j}e^{-i\omega_{j} t} - v_{j}^{*}e^{i \omega_{j} t}|^{2}$, where $\psi_0$ indicates the ground state wave function, shows a standing wave with $n_x$ and $n_y$ nodes along each axis.
In Fig. \ref{spectrum}, the quantum numbers $(n_x, n_y)$ are indicated for each line, and it is noted that different slopes correspond to different $n_y$, as we are using the $\omega_x$ trap dimension.
A number of avoided crossings between spectral lines are clearly observed in Fig. \ref{spectrum}.
Since the ``perturbation" added to the Hamiltonian due to the aspect ratio change, $M(x^2\omega_x\delta\omega_x+y^2\omega_y\delta\omega_y)$, involves even parity, these avoided crossings are found to occur only when two non-crossing states have the same parity \cite{LandauQuantum}. 
It should be noted that avoided crossings are found regardless of the dimensionality of the system \cite{Reidl}.
For high $\lambda$, in the low energy regime, $n_y=0$ modes are decoupled from the others to form a group of 1D excitations, for which the wavelength of the excitation along the $x$ axis is much larger than the size of BEC along the $y$ axis.  

\begin{figure}[ht]
\centerline{(a)\ $\lambda = 3.1$\hspace{1cm}(b)\ $\lambda = 3.3$\hspace{1cm}(c)\ $\lambda = 3.5$}
\centerline{\includegraphics*[height=2.5cm]{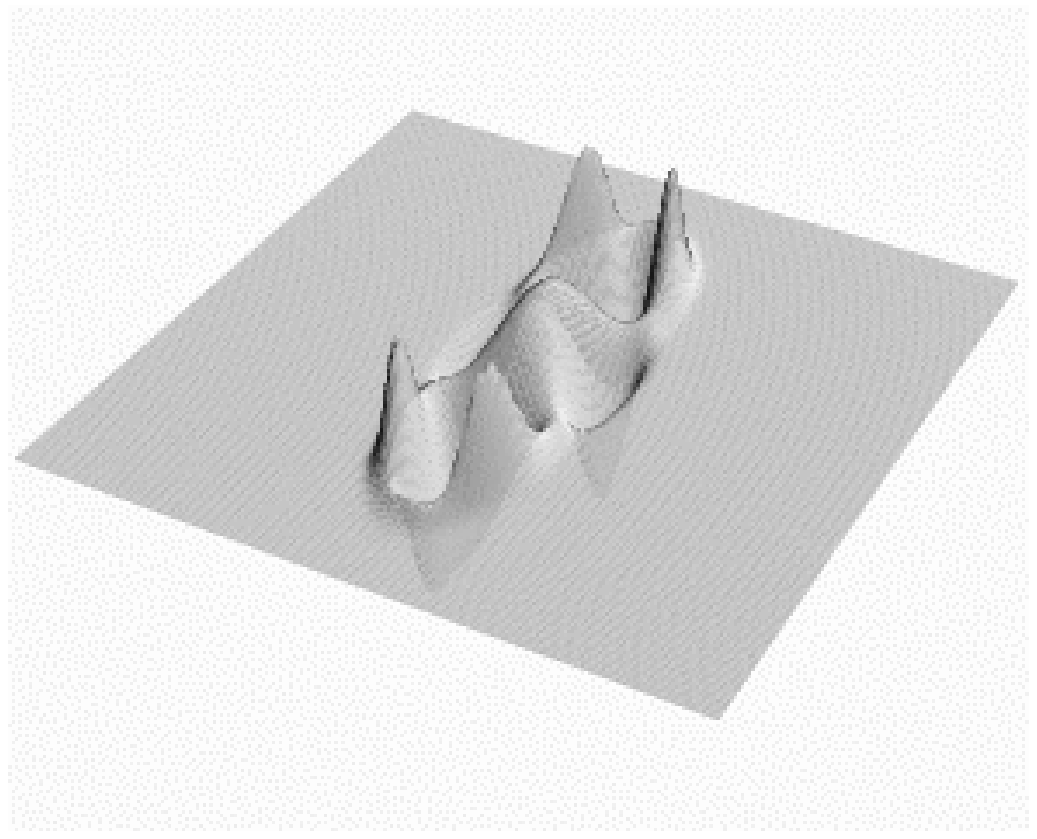}\hspace{-.5cm} 
\includegraphics*[height=2.5cm]{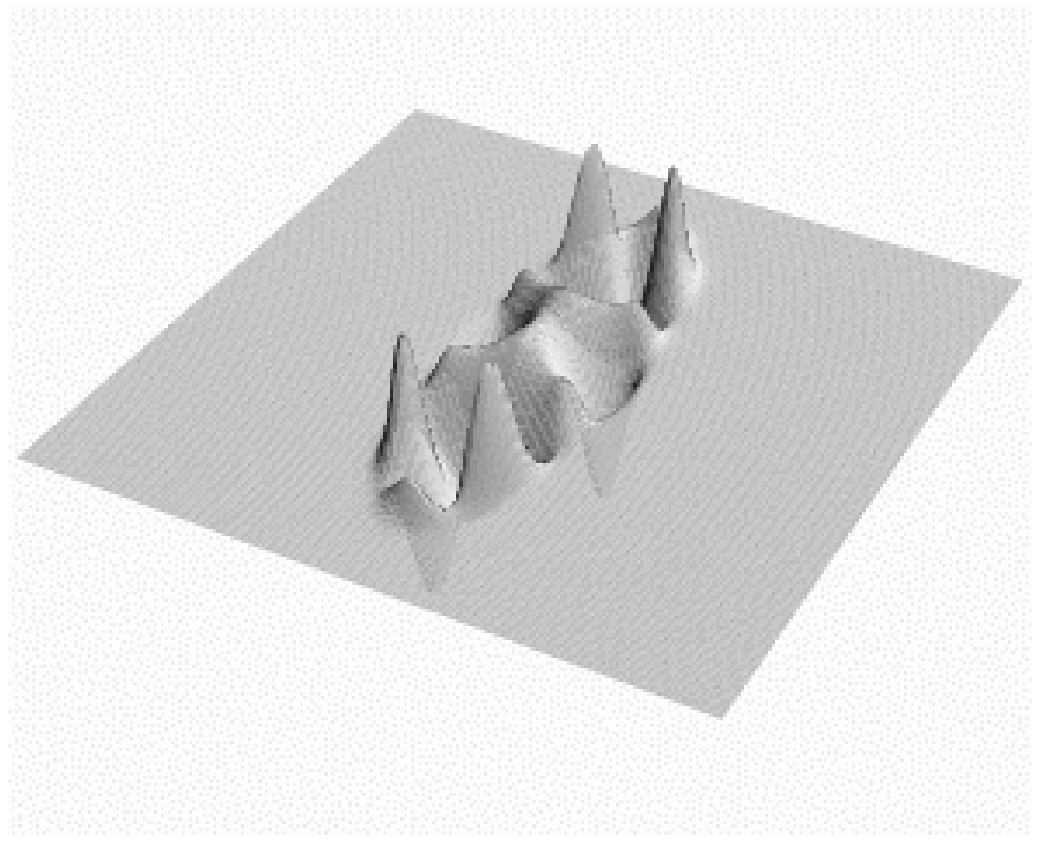}\hspace{-.5cm} 
\includegraphics*[height=2.5cm]{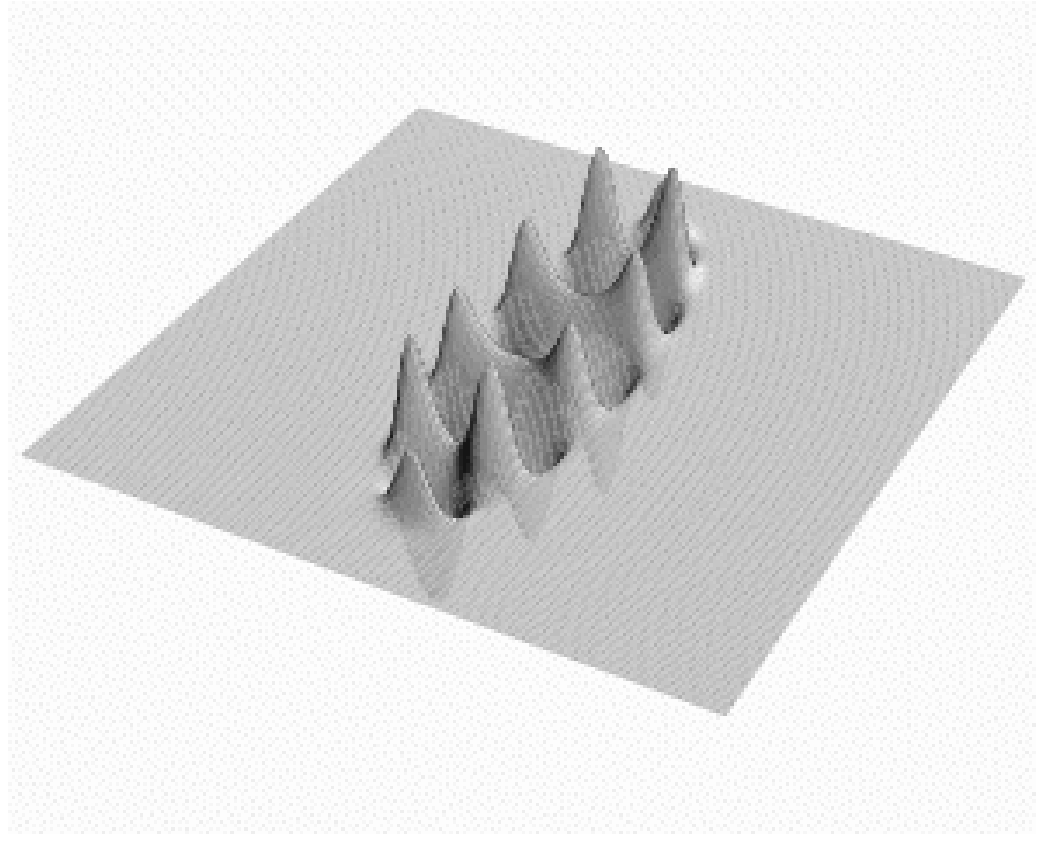}}
\vspace{-.2cm}
\caption{Change in the density fluctuation $n'$ for the energy eigenstates along the lower spectral line associated with the avoided crossing in the circle drawn into Fig. \ref{spectrum}.}
\label{avoided crossing}
\end{figure}
We focus our attention on a specific avoided crossing between states $(n_x =2,n_y =2)$ and $(n_x =10,n_y =0)$ which is circled in Fig. \ref{spectrum}.
Figure \ref{avoided crossing} shows the change of the function $n'_j\equiv\psi^*u_j-\psi v_j$ \cite{DalfovoReview} for an energy eigenstates along the lower spectral line of the avoided crossing.
Figure \ref{avoided crossing} (a) and (c) correspond to $(2,2)$ and $(10,0)$ modes with $\lambda=3.1$ and $3.5$ respectively.
We shall refer to the asymptotic states far from the center of the avoided crossing such as (a) or (c) ``shape eigenstates" as they have definite number of zeros along each axis.
The energy eigenstate at the avoided crossing, Fig. \ref{avoided crossing} (b), is a linear combination of the two shape eigenstates, (a) and (c), and is itself not a shape eigenstate.

If the system undergoes an adiabatic change in its aspect ratio around the avoided crossing starting from one of the asymptotic states, it cannot go through the energy gap between the two avoided spectral lines \cite{Landau-Zener}, resulting in the exchange of the mode characteristics;
for instance, starting from a $(2,2)$ mode, it ends up in $(10,0)$ mode.
For a diabatic transition, the state may tunnel through the energy gap between the two avoided spectral lines to remain in the same shape eigenstate.  
According to the LZ theory \cite{Landau-Zener}, the probability for the state to remain in the same shape eigenstate is $P = e^ {-2\pi\gamma}$, where $\gamma = \hbar^{-1}(\Delta /2)^2\left[(d/dt)(\epsilon_1 - \epsilon_2)\right]^{-1}$.
Here, $\epsilon_1$ and $\epsilon_2$ are the eigenenergies corresponding to two avoided states and $\Delta$ is the minimum value for $|\epsilon_1-\epsilon_2|$.
For the above avoided crossing $\Delta \approx 0.07$ and $(d/dt)(\epsilon_1-\epsilon_2) = \dot{\lambda} (d/d\lambda)(\epsilon_1-\epsilon_2) \approx 1.4\dot{\lambda}$, which gives $P=e^{-5.5\times 10^{-3}\dot{\lambda}^{-1}}$ in the $\omega_x$ trap dimension.
Here, $\dot{\lambda}$ denotes $d\lambda/dt$ and $\dot{\lambda} = 0.001$ corresponds to the adiabatic case with $P=0.004$, while $\dot{\lambda} = 0.1$ to the diabatic case with $P=0.95$.
In order to observe this adiabatic transition experimentally, the change in $\lambda$ from 3.1 to 3.5 should happen throughout the time longer than $\Delta t = \Delta\lambda/\dot{\lambda} = 400 \approx 64\times 2\pi \omega_x^{-1}$.

We have numerically simulated the temporal evolution of a chosen excitation with $\lambda$ varying in time by using time-dependent GPE.
As an initial state, we chose the normal mode $(2,2)$ for $\lambda=3.1$. 
To visualize more clearly the excitation possessed by the condensate as it evolves, we subtracted the ground state from the total wavefunction to give time-dependent density fluctuation $\delta\rho(t)\equiv\psi^*(t)\psi(t)-\psi^*_0\psi_0$. 
Assuming that there are $|q_j|^2$ quasiparticles for each excitation mode $j$, as far as $|q_j|^2 \ll N$, the time evolution of the mean field is almost exactly given by \cite{DalfovoReview}
\begin{equation}
\psi(t)=\psi_0+\sum_j\left(q_ju_j e^{-i\omega_jt}-q_j^*v_j^* e^{i\omega_jt}\right),
\label{t-evol of psi}
\end{equation}
which, with real $u_j$, $v_j$, $\psi_0$ and $q_j$, yields 
\begin{equation}
\delta\rho(t)\approx\sum_j 2q_j n'_j\cos(\omega_jt),
\label{delta_rho}
\end{equation}
with higher order terms of $q_j$ ignored. 
Noting that each term in $\delta \rho (t)$ is a product of $n'_j$ with a sinusoidal time dependence $\cos (\omega_j t)$, we have compared $\delta \rho (t)$ with the exact energy eigenstates given in Fig. \ref{avoided crossing}.
We have confirmed that the shape of $\delta\rho(t)$ changes from Fig. \ref{avoided crossing}(a) to (c) under the adiabatic change in $\lambda$, while the state remains in the same shape eigenstate for the diabatic case. 

\begin{figure}
\centerline{\includegraphics*[height=3cm]{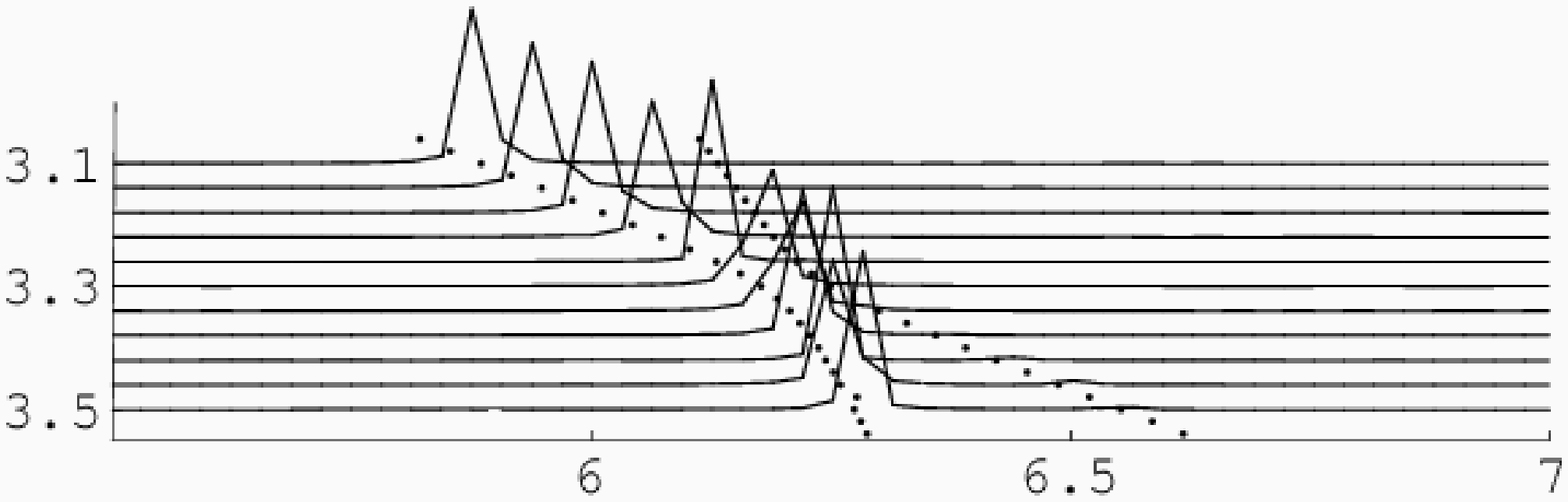}}
\vspace{-.2cm}
\centerline{(a)}
\centerline{\includegraphics*[height=3cm]{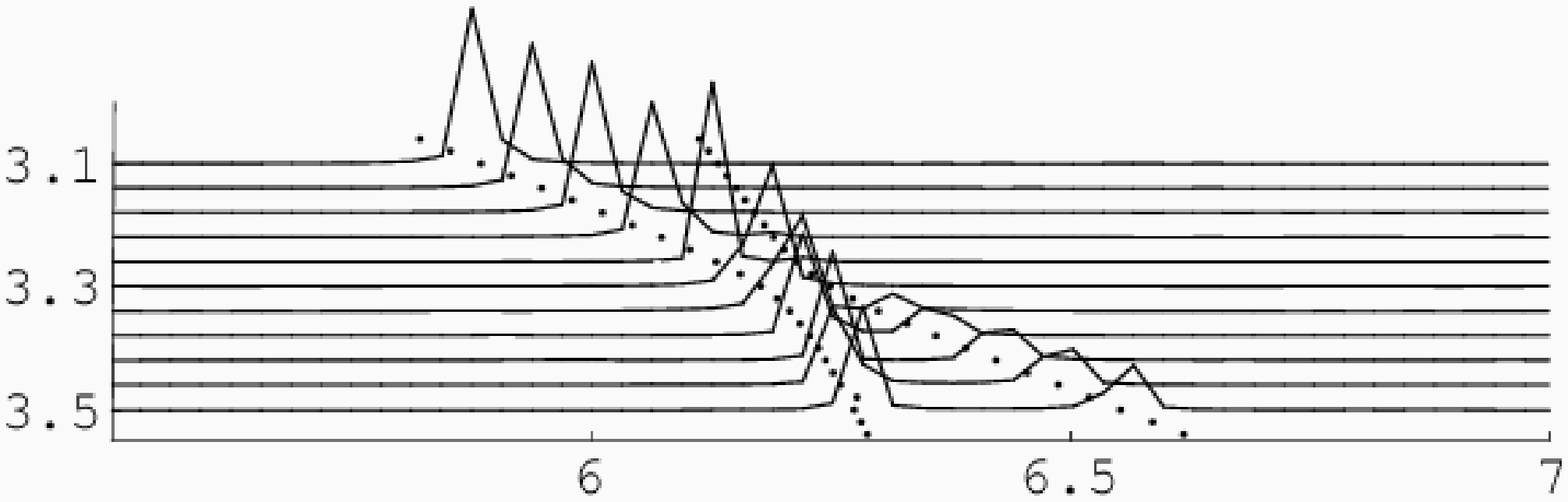}}
\vspace{-.2cm}
\centerline{(b)}
\centerline{\includegraphics*[height=3cm]{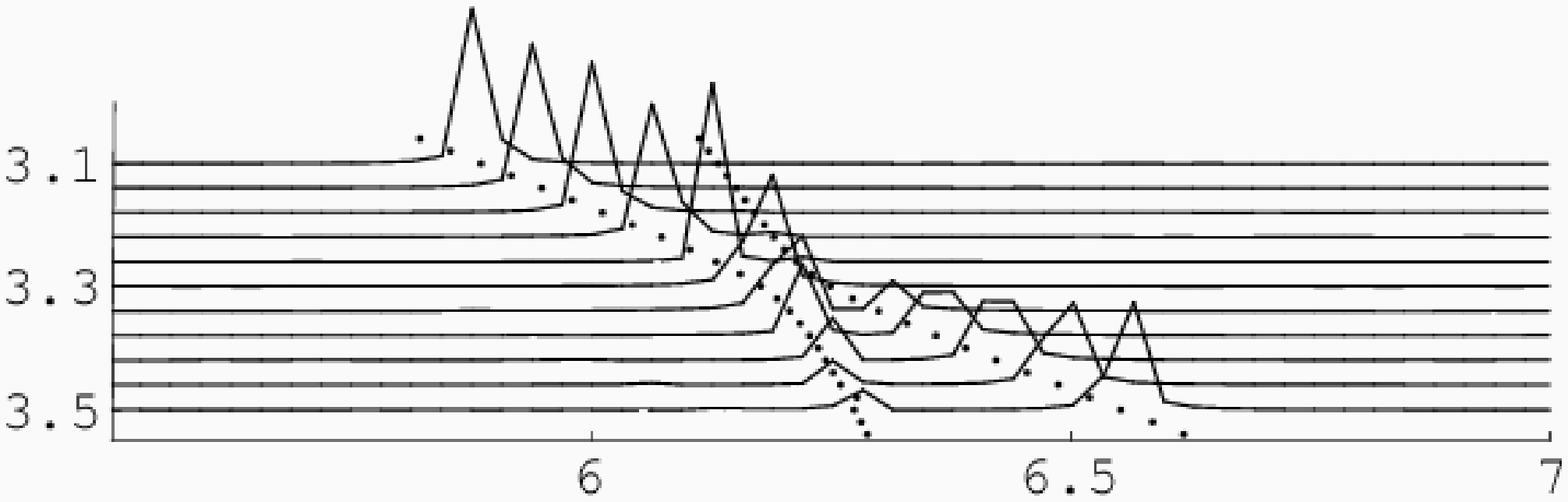}}
\vspace{-.2cm}
\centerline{(c)}
\caption{Distribution function in the frequency space for each value of $\lambda$ while $\lambda$ increases from $3.1$ to $3.5$ (a) adiabatically ($\dot{\lambda}=0.001$), and (c) diabatically ($\dot{\lambda}=0.1$).  (b) is for the intermediate case ($\dot{\lambda}=0.01$). 
}
\label{spectral_analy}
\end{figure}

In order to estimate the amount of transition quantitatively, we have used spectral analysis. 
Equation (\ref{delta_rho}) shows that the quasiparticle population for each energy mode can be obtained by performing a time-domain Fourier transformation of $\delta\rho(t)$.
As $\lambda$ changes in time the quasiparticle populations change as a function of $\lambda$.
Figure \ref{spectral_analy} shows the distribution function in the frequency space for each $\lambda$.
For the adiabatic case (a), the peaks nicely follow a single spectral line of two avoided energy levels, while for the diabatic case (c), the population peak tunnels through from one spectral line into the other to stay in the same shape eigenstate.
In addition, we obtained the result for the intermediate case (b) at $\dot{\lambda}=0.01$, which shows the splitting of the initial population on one energy eigenstate into two after going through the avoided crossing region.
It is notable that, although these simulations were carried out using full GPE, the linear LZ theory is still valid with the quasiparticle populations small enough compared to the ground state population.

We believe that the rate dependence of the process may have an implication in the process of releasing condensates with specific excitations from a trapping potential, which is usually done in the experiments. 
In addition, the roughness of the potential in the atomic waveguide could introduce the change of aspect ratio naturally, in which case, the reaction of the phonon modes may have a strong dependence on the velocity of a moving condensate.

\begin{figure}
\centerline{(a)\hspace{3.5cm}(b)}
\centerline{\includegraphics*[height=4cm,width=4.5cm]{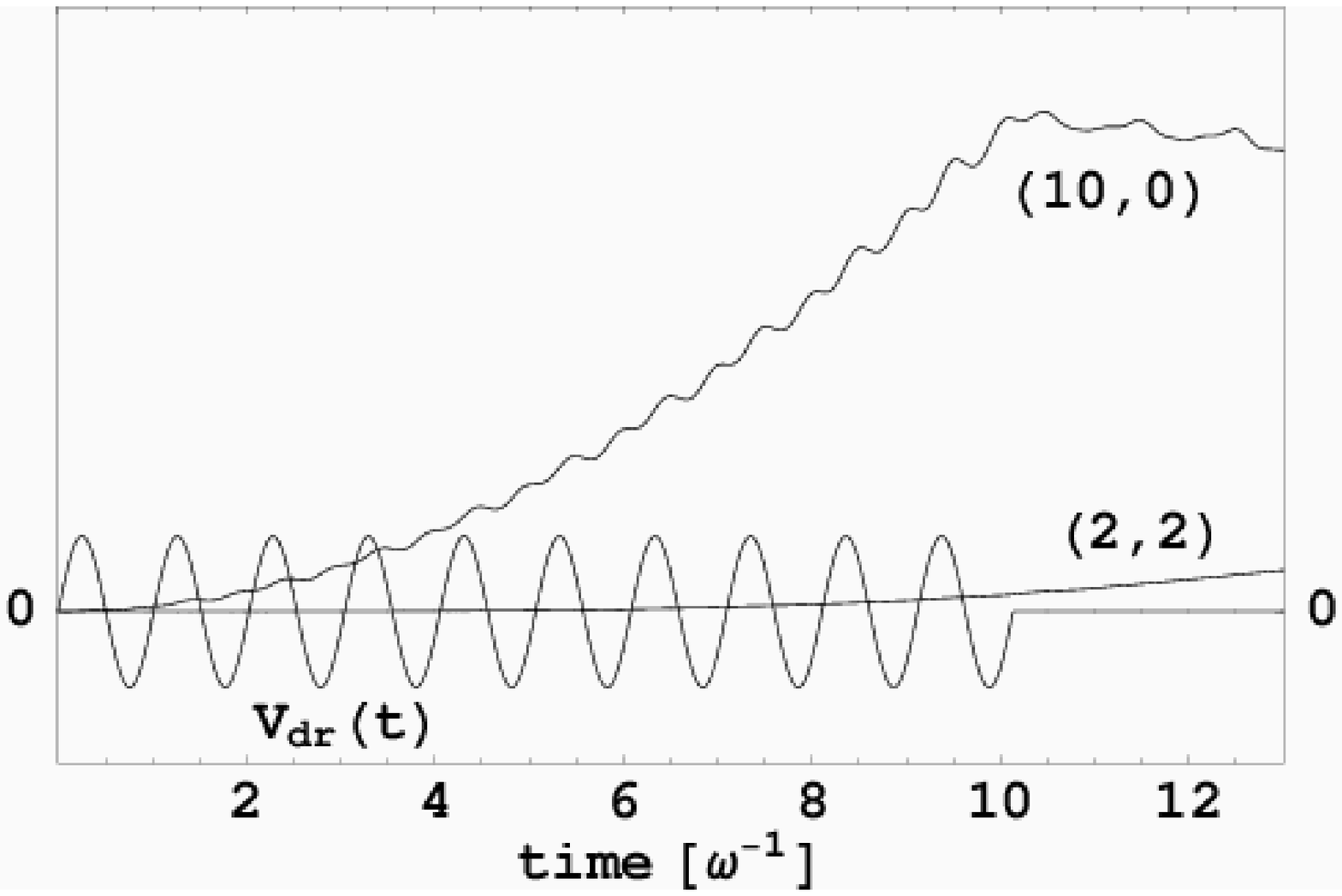}
\hspace{0cm}\includegraphics*[height=4cm,width=4.5cm]{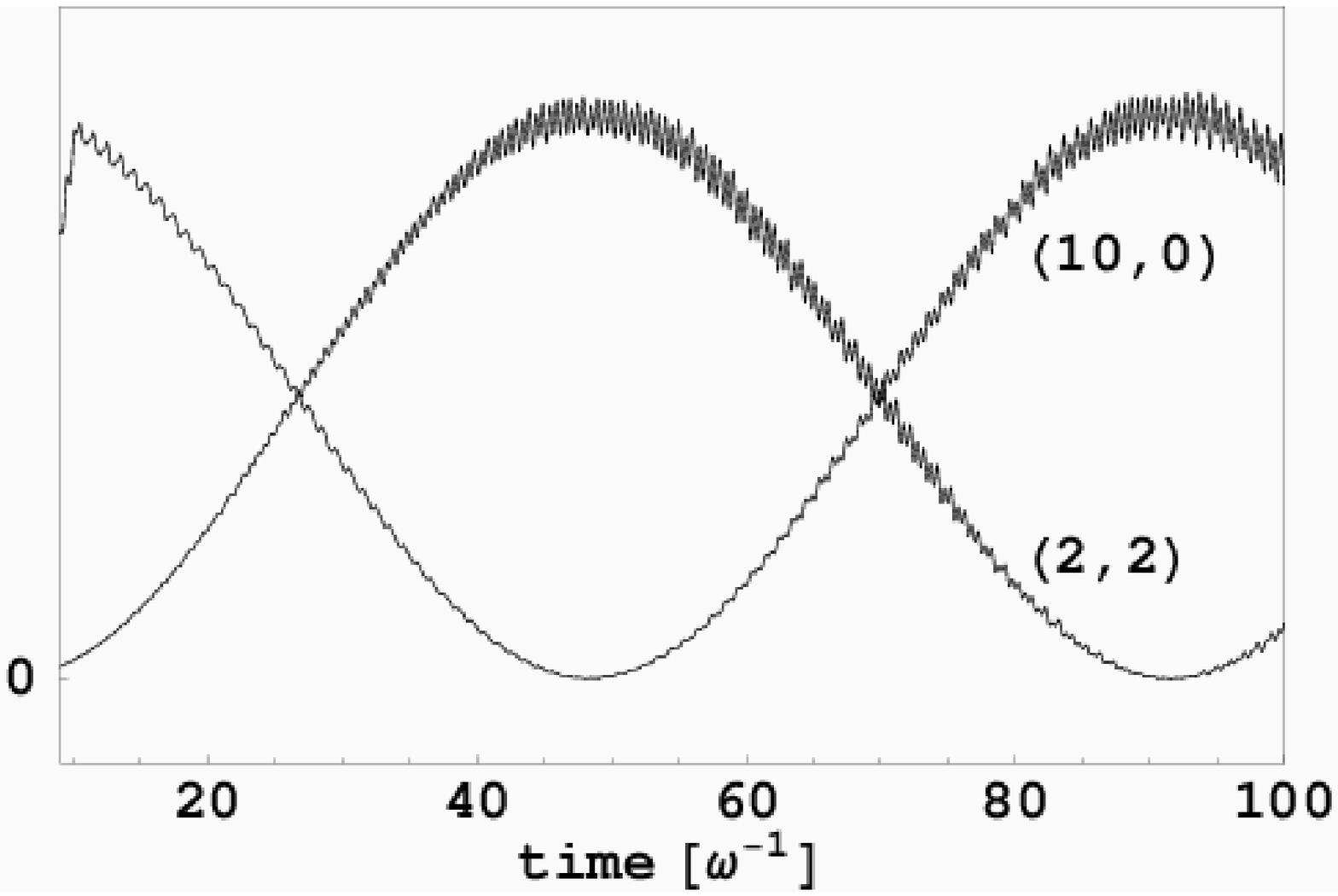}}
\caption{(a) The population growth of the $(10,0)$ mode while the ground state of BEC is driven by $V_{\rm dr}(t)\sim\sin (\omega t)$. The other mode $(2,2)$ is clearly suppressed, and (b) the follow-up population oscillation between $(10,0)$ and $(2,2)$ modes near the avoided level crossing.}
\label{shape oscil}
\end{figure}

We now consider an additional scenario for quantum control.
With a correct driving potential it is possible to excite energy eigenstates or certain shape eigenstates.
However, if the generated shape eigenstate coincides with an avoided crossing for some specific aspect ratio $\lambda$, the excitation is not an energy eigenstate but a linear combination of two almost-degenerate energy eigenstates.
If one denotes the two energy eigenstates as $\chi_i$ and the two orthogonal shape eigenstates as $\phi_i$, these two different basis sets are related by a unitary transformation $U_{ij}$.
The temporal evolution of an initial shape eigenstate, $\psi(t=0)\equiv\phi_1=U_{11}\chi_1+U_{12}\chi_2$, is then $\psi(t)=U_{11}\chi_1e^{-i\omega_1t}+U_{12}\chi_2e^{-i\omega_2t}$, such that the population of mode 
$\phi_1$ is $\left|\langle\phi_1|\psi(t)\rangle\right|^2=|U_{11}|^4+|U_{12}|^4 +2|U_{11}|^2|U_{12}|^2\cos(\omega_1-\omega_2)t$,
while $\left|\langle\phi_2|\psi(t)\rangle\right|^2=1-\left|\left<\phi_1|\psi(t)\right>\right|^2$.
Clearly the populations oscillate between two modes even though it starts from only one of the two states.  
In our numerical simulations, we could efficiently generate a shape eigenstate with a specific quantum number $(n_x, n_y)$ by applying a driving external potential $V_{\rm dr}(t)$ to the trap potential $V_{\rm tr}$ of the form:
\begin{equation}
V_{\rm dr}(t) \sim \sin(\omega t)\cos\left(\frac{2\pi x}{\lambda_x}\right)\cos\left(\frac{2\pi y}{\lambda_y}\right),
\label{V_dr}
\end{equation}
where $\omega$, $\lambda_x$ and $\lambda_y$ are the frequency and the estimated wavelengths for the aimed normal mode respectively.
A $(10,0)$ mode at $\lambda=3.3$ was generated using $V_{\rm dr}(t)$ with $\omega=6.2$, $\lambda_x\approx 3$ and $\lambda_y\rightarrow\infty$. 
When $V_{\rm dr}$ was turned off, the oscillation between the generated state and the $(2,2)$ mode was clearly observed in the evolving density fluctuation $\rho(t)$.
In order to show this ``shape oscillation" quantitatively, we adopted quasiparticle projection method \cite{Morgan}, in which the amplitude of the $j$th mode from Eq. (\ref{t-evol of psi}) is  $q_j=\int d{\bf r} (\tilde{u}_j^*\psi(t) e^{i\mu t} + \tilde{v}_j^*\psi^*(t)e^{-i\mu t})$, where $\tilde{u}_j=u_j-a_j\psi_0/N_0$, $\tilde{v}_j^*=v_j^*-a_j^*\psi_0/N_0$ and $a_j=\int d{\bf r}\psi_0^*u_j=\int d{\bf r}\psi_0v_j$.
Fig. \ref{shape oscil}(a) shows the population growth of the $(10,0)$ mode in the BEC driven by $V_{\rm dr}(t)$, while the population of the $(2,2)$ mode is suppressed. 
Fig. \ref{shape oscil}(b) represents the follow-up population oscillation between two shape eigenstates after turning off the driving potential. 
The period of the shape oscillation matches with the expected $2\pi/(\omega_1-\omega_2)\approx 90$.
A related result was reported in Ref. \cite{Dalfovo} which was discussed in a different context of resonant nonlinear mode coupling. 
It is noted that a similar oscillation can also be found in particle physics, in which neutrinos, generated in a muon or electron eigenstate, are slightly rotated from the exact energy eigenstate and undergo what is known as ``neutrino oscillation" \cite{Kuo}.

\begin{figure}
\centerline{\includegraphics*[width=7cm]{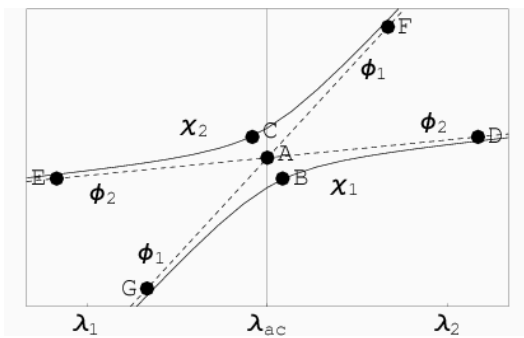}}
\caption{Schematic diagram of an avoided crossing. $\phi_i$ indicate asymptotic shape eigenstates and $\chi_i$ exact energy eigenstates.
}
\label{avoided crossing2}
\end{figure}

Finally, we would like to mention further possible quantum control of BEC involving phonon modes around avoided crossing. 
We have observed that the transition between two different shape eigenstates happens under adiabatic process. 
The same kind of transition, however, may also be made with a different process (See Fig. \ref{avoided crossing2}):
Starting with a shape eigenstate $\phi_1$ at $\lambda=\lambda_1$, make a diabatic or sudden change of $\lambda$ to $\lambda_{ac}$ which does not change the mode, wait until the shape oscillation achieves $\phi_2$, and then make a diabatic change of $\lambda$ again from $\lambda_{ac}$ to $\lambda_2$ ending up with the other shape eigenstate $\phi_2$ at $\lambda_2$ ($G\rightarrow A\rightarrow D$).
If we do the second diabatic process backward from $\lambda=\lambda_{ac}$ to $\lambda_1$, we are able to get mode transition from $\phi_1$ to $\phi_2$ at the same $\lambda =\lambda_1$ only by changing the aspect ratio slightly back and forth ($G\rightarrow A\rightarrow E$).
Another possible process is to change $\lambda$ from $\lambda_1$ to $\lambda_2$ and back to $\lambda_1$ with a fixed $|\dot{\lambda}|$ and with the shape eigenstate $\phi_1$ initially. 
The final state will end up with the same shape eigenstate for both diabatic ($G\rightarrow A\rightarrow F\rightarrow A\rightarrow G$) and adiabatic ($G\rightarrow B\rightarrow D\rightarrow B\rightarrow G$) case, except for the intermediate case, in which the final state will have considerable population on both shape eigenstates. This process may be compared to the collision of the second kind in an atomic collision \cite{LandauQuantum}.  
These could be useful if the BEC phonon modes could ever be used in the future as qubits for quantum computation.

In summary, we have studied ways to quantum control quasiparticles in BEC with the aspect ratio of the harmonic trapping potential as a parameter. 
Using numerical simulations, we confirmed that LZ transition does occur in the linearized regime. 
We have also observed that a shape oscillation between different types of collective excitations are generated at a specific trap anisotropy.
This ``beating" observed in the condensate density may also be viewed as a signature for the existence of an avoided crossing in the Bogoliubov spectrum.
These phenomena should be readily observable experimentally.
Future work would involve extension of this work to a rotating BEC with vortex lattices \cite{Oktel}. 
This work is supported by the NSF, DOE, ONR and ARO.

\end{document}